\pdfoutput=1
\documentclass[12pt]{article}

\usepackage{graphicx,psfrag,epsf}
\usepackage{enumerate}
\usepackage{natbib}
\usepackage{url} % not crucial - just used below for the URL 

\newcommand{\blind}{1}

\usepackage{amsmath}
\usepackage{amssymb}
\usepackage{amsthm}
\usepackage{mathtools}

\usepackage{array,multirow}
\usepackage{floatrow}
\usepackage{caption}
\usepackage{rotating}
\usepackage{tabu}

\usepackage{bm}
\usepackage{bbm}

\theoremstyle{plain}
\newtheorem{proposition}{Proposition}[section]

\begin{document}

\def\spacingset#1{\renewcommand{\baselinestretch}%
{#1}\small\normalsize} \spacingset{1}

%%%%%%%%%%%%%%%%%%%%%%%%%%%%%%%%%%%%%%%%%%%%%%%%%%%%%%%%%%%%%%%%%%%%%%%%%%%%%%

\if1\blind
{
  \title{\bf Convergence diagnostics for MCMC draws of a
categorical variable}
  \author{Deonovic, Benjamin E.\\
    Department of Biostatistics, University of Iowa\\
    and \\
    Smith, Brian J. \\
    Department of Biostatistics, University of Iowa}
  \maketitle
} \fi

\if0\blind
{
  \bigskip
  \bigskip
  \bigskip
  \begin{center}
    {\LARGE\bf Convergence diagnostics for MCMC draws of a
categorical variable}
\end{center}
  \medskip
} \fi

\bigskip
\begin{abstract}
Markov Chain Monte Carlo (MCMC) is a popular class of statistical methods for simulating autocorrelated draws from target distributions, including posterior distributions in Bayesian analysis. An important consideration in using simulated MCMC draws for inference is that the sampling algorithm has converged to the distribution of interest. Since the distribution is typically of a non-standard form, convergence cannot generally be proven and, instead, is assessed with convergence diagnostics. Although parameters used in the MCMC framework are typically continuous, there are many situations in which simulating a categorical variable is desired. Examples include indicators for model inclusion in Bayesian variable selection and latent categorical component variables in mixture modeling. Traditional convergence diagnostics are designed for continuous variables and may be inappropriate for categorical variables. In this paper two convergence diagnostic methods are considered which are appropriate for MCMC data. The diagnostics discussed in the paper utilize chi-squared test statistics for dependent data. Performance of the convergence diagnostics is evaluated under various simulations. Finally, the diagnostics are applied to a real data set where reversible jump MCMC is used to sample from a finite mixture model.
\end{abstract}

\noindent%
{\it Keywords:}   Time series, chi-squared, discrete, dependent
\vfill

\newpage
\spacingset{1.45} % DON'T change the spacing!

\section{Introduction}
In Bayesian statistics, the recent development of MCMC methods and the increase in computational resources have provided statisticians the ability to sample from complex posterior distributions, which often require the integration of many unknown parameters. As an MCMC sampling algorithm proceeds, the distribution from which the samples are being drawn converges towards a target distribution. In all but the most trivial examples, this convergence cannot be proven, but rather must be empirically tested using convergence diagnostics. Convergence diagnostics play an integral part in assessing the reliability of parameter summaries in MCMC output. The goal of MCMC is often not only to draw samples from some distribution, but to do inference on that distribution. Therefore, reliable summary quantities are paramount. 

Although many convergence diagnostics have been developed, the assumptions of these diagnostics are not amenable to models with discrete parameters, which are becoming popular. For a review of classical MCMC convergence diagnostics see \cite{cowles:1996:diag}. When the parameter of interest in MCMC is discrete, key assumptions of these diagnostics and their tests are either violated or require large samples. The rise in popularity of models with discrete parameters can be seen in the large number of application areas including change-point models, where the number and location of change points are unknown; finite mixture models, where the number of mixing components are unknown; variable selection, where the parameters to include are unknown; and Bayesian nonparametrics, where the location and number of knots are unknown. As a result of the increase in popularity of discrete parameters in MCMC, convergence diagnostics that do not impose burdensome assumptions for discrete parameters need to be developed.

\subsection{Other methods}
\label{sec:other}
The development of transdimensional MCMC, such as the reversible jump MCMC (RJMCMC) sampler by \cite{green:1995:rjmcmc}, has been a primary reason for the spike in interest for models with discrete parameters. In transdimensional MCMC, the continuous parameters of interest, say $\bm\theta$, may vary in dimension at each step of the MCMC algorithm. By construction, models sampled by transdimensional MCMC have discrete parameters. For example, the dimension of $\bm\theta$ can be represented by a discrete parameter or, if the dimension of $\bm\theta$ varies due to a subset of $\bm\theta$ being selected at each iteration, an indicator of which elements of $\bm\theta$ are included in the current iteration, can be thought of as a discrete parameter. 

Due to the popularity of the sampler by Green, several MCMC diagnostics have been developed specifically for output from transdimensional MCMC. \cite{fan:2011:rjmcmc} provide a review of RJMCMC along with the associated MCMC diagnostics. Since these diagnostics could be used to assess convergence of models with discrete parameters, a brief overview of them is provided. 

In \cite{brooks:2003:NPC}, between-chain convergence is assessed using nonparametric hypothesis tests such as Pearson's chi-squared and Kolmogorov-Smirnov tests. Since these tests require independent data they are not appropriate for output from MCMC. To overcome this limitation, Brooks et al.\ require the MCMC output to be heavily thinned (only retain every $m$th iteration, for large $m$) in order to reduce the autocorrelation in the output. Thinning the MCMC output as a method of reducing the autocorrelation is not desirable as this reduces the number of samples to approximate the posterior distribution for inference. 

\cite{gelman:1992:IIS} utilize an ANOVA approach to compare the variance of a continuous parameter both between and within chains. \cite{brooks:2000:ANOVA} expand this approach to two-way ANOVA by including the discrete parameter as a factor in the model. Their approach is designed to monitor some function of continuous parameters $\bm\theta$ involved in the model. Thus, the user must identify some function of the parameters that retains its interpretation as the dimensions of $\bm\theta$ change. The authors suggest the deviance as a default. \cite{castelloe:2002:CARJ} further extend this idea to an unbalanced (weighted) two-way ANOVA to prevent the statistics from being dominated by a few visits to rare models. They also allow the user to track multiple functions rather than just one. Ultimately, all of these ANOVA based methods assess the convergence of transdimensional MCMC by focusing on the continuous parameters. Although this may be appropriate when using transdimensional MCMC sampling algorithms, if a separate algorithm is used for the discrete parameters a more direct diagnostic that focuses on the discrete parameter would be preferred. 

\cite{sisson:2007:conv} construct a diagnostic for models that can be formulated as marked point processes. Let $\mathbf{v}$ be user-selected reference points for the continuous parameter $\bm\theta$ and $\bm{\theta}^{(j)}_1,\ldots,\bm{\theta}_N^{(j)}$ MCMC draws of the posterior of $\bm\theta$ where $\bm{\theta}_i = (\theta_{i1},\ldots,\theta_{ik_i})^\intercal$ is a $k_i$ dimensional vector and $j=1,\ldots,c$ indexes $c$ independent chains. For each $v\in\mathbf{v}$ let $x^{(i)}$ be the distance from $v$ to the nearest component of $\bm{\theta}_i$. The empirical distribution function is then estimated for these distances for each chain, $\hat{F}^{(j)}(x,v) = N^{-1}\sum_{i=1}^N \mathbbm{1}\left\{ x^{(i)} \leq x\right\}$ for $j=1,\ldots,c$. Using $L^p$ distance with $p\in \mathbb{R}^+$ as the difference between these empirical distributions provides a measure of similarity for two chains: 
\[
|\mathbf{v}|^{-1}\sum_{v\in\mathbf{v}} \int_0^\infty |\hat{F}^{(i)}(x,v) - \hat{F}^{(j)}(x,v)|^p dx
\]

Although these diagnostics perform well in the contexts for which they were designed, they are unsatisfactory as general MCMC diagnostics for discrete parameters. These diagnostics either can only be used on output from the RJMCMC sampler, rather than any discrete parameter MCMC output; make non-optimal assumptions; or require additional user-specified input. Therefore an MCMC diagnostic that is constructed for a discrete state space but overcomes the shortcomings of the diagnostics developed for the RJMCMC sampler output would be a boon. 

In this paper two such diagnostics are developed. Section \ref{sec:meth} describes a general approach for assessing between-chain and within-chain convergence. In Section \ref{sec:met1} a chi-squared test statistic is constructed based on comparing the frequency distribution of the categories of the discrete parameter. Section \ref{sec:met2} provides an alternative chi-squared test statistic using the estimated transition matrices. A simulation study is conducted in Section \ref{sec:sim} to evaluate the power and type I error of these methods. Section \ref{sec:data} compares the two methods in this paper to the method from \cite{sisson:2007:conv} on a real data set. 

\section{Methods}
\label{sec:meth}
Two MCMC diagnostics were developed to assess convergence of posterior draws from MCMC sampler output on a discrete state space. Both of these methods diagnose the similarity of two independent portions of the output. Therefore, both methods can be adapted to assess the convergence of MCMC draws within a chain and between multiple chains. To assess convergence within a chain an approach similar to the one taken in \cite{geweke:1992:EAS} is used. In particular, a specified portion (e.g. 35\%) of the beginning of the chain is compared to some portion of the end of the chain. The diagnostics require these two portions to be independent, so there needs to be space between the beginning and ending portions. To assess convergence between chains, the chains are simply taken as the independent units to be compared. 

\subsection{Method 1: Frequency Distribution}
\label{sec:met1}
\begin{table}[ht!]
  \centering
  \begin{tabular}{cc|ccc|cc}
    & \multicolumn{1}{c}{} & \multicolumn{3}{c}{Segment} & \multicolumn{1}{c}{}& \multicolumn{1}{c}{}\\
    & \multicolumn{1}{c}{} & \multicolumn{1}{c}{1} & \multicolumn{1}{c}{$\hdots$}   &  \multicolumn{1}{c}{$s$}  & \multicolumn{1}{c}{}& \multicolumn{1}{c}{}\\
    \cline{3-5}
    \parbox[t]{2mm}{\multirow{3}{*}{\rotatebox[origin=c]{90}{Categories}}} &  1 & $N_{1}^{(1)}$ & $\hdots$ & $N_{1}^{(s)}$ & $N_{1}$\\
    %\cline{2-4}
                                                                           &  $\vdots$ & $\vdots$ & $\ddots$ & $\vdots$ & \\
    %\cline{2-4}
                                                                           &  $r$ & $N_{r}^{(1)}$ & $\hdots$ & $N_{r}^{(s)}$ & $N_{r}$\\
    \cline{3-5}
    & \multicolumn{1}{c}{} & \multicolumn{1}{c}{$n_1$} & \multicolumn{1}{c}{$\hdots$}   &  \multicolumn{1}{c}{$n_s$}  & \multicolumn{1}{c}{$n$}&
  \end{tabular}
  \caption{Tabularized Markov Chain data for Categorical Convergence Test. Data from $s$ Markov chains tabularized by category and chain.}
  \label{tab:chpt2t1}
\end{table}
Method 1 aims to test the similarity of the frequency distribution of the discrete parameter between independent segments, similar to Pearson's chi-squared statistic. This section introduces a test statistic that measures the discrepancy of the frequency distribution between independent segments and describes four procedures for determining whether this discrepancy is significant. 

\subsubsection{Hangartner Procedure}

Let $X^{(1)}_t, \ldots, X^{(s)}_t$ be $s$ independent, categorical time series of length $n_i$ for $i=1,\ldots,s$ respectively, which take on values in $\mathcal{V}=\left\{1,\ldots,r\right\}$. Due to the focus on the development of convergence assessments for output from MCMC, assume further that these time series are reversible Markov chains. As mentioned above, these time series could either be the independent chains from an MCMC run or segments from a single MCMC chain that are separated by enough iterations so as to be considered independent. Let $Y^{(i)}_{tj}$ be the binarization of $X^{(i)}_t$ such that 
\[
Y^{(i)}_{tj} = \begin{cases}
1 & \text{if $X^{(i)}_t = j$}\\
0 & \text{otherwise}
\end{cases}
\]
Let $N_j^{(i)}(T)$ be the number of iterations up to iteration $T$ that take on the value $j$ in the $i$th segment, i.e.\ $N_j^{(i)}(T) = \sum_{t=1}^T Y_{tj}^{(i)}$ where $i=1,\ldots,s$ and $j=1,\ldots,r$. For conciseness let $N_j^{(i)} = N_j^{(i)}(n_i)$. Such data is often displayed in tabular format as in \ref{tab:chpt2t1}. The standard chi-squared test of homogeneity is 
\begin{equation} 
X^2 = \sum_{i=1}^s \sum_{j\in R} \dfrac{n_i(\hat{p}_j^{(i)} - \hat{p}_j)^2}{\hat{p}_j}
\label{chpt2eq1}
\end{equation}
where $\hat{p}_j^{(i)} = N_j^{(i)}/n_i$ is the estimated proportion of category $j$ in segment $i$, $N_j = \sum_{i=1}^s N_j^{(i)}$ is the total number of transitions from state $j$, $\hat{p}_j=\sum_{i=1}^s N_j^{(i)} / \sum_{i=1}^s n_i$ is the estimated proportion of category $j$ by pooling the segments together, and $R = \left\{j|N_j > 0\right\}$. If each iteration in the categorical time series were independent from one another then $X^2$ would have a $\chi^2$ distribution with $(|R| -1)(s-1)$ degrees of freedom \citep{cramer:1946:mms}. 
The $X^2$ test statistic utilizing the asymptotic distribution as an indicator of significant discrepancy was proposed by \cite{hangartner:2011:disc} as an MCMC diagnostic for discrete state space parameters. The diagnostic has several benefits: it does not rely upon the estimation of spectral density (such as \cite{geweke:1992:EAS} or \cite{heidelberger:1983:SRL}), on suspect normality assumptions (such as \cite{gelman:1992:IIS}), or on determining overdispersion within a small number of outcomes (such as \cite{gelman:1992:IIS}), all of which can be problematic with discrete measures. However, since the draws from an MCMC sampler are not independent, the test statistic will be overly liberal in identifying differences between segments because it does not account for the autocorrelation. 

\subsubsection{Wei\ss\ Procedure}
The Pearson chi-squared test statistic needs to be adjusted to account for autocorrelation when the data are not independent, such as data from MCMC. To make such an adjustment, assume the data follow an NDARMA($p,q$) model described by \cite{jacobs:1983:NDARMA} (see supplementary materials). Let $X_t$ be a categorical time series which follows an NDARMA model. An important quantity for NDARMA models is 
\begin{equation}
c = 1 + 2\sum_{t=1}^\infty \text{corr}(X_1,X_{1+t}). \label{chpt2eqc}
\end{equation} \cite{weiss:2008:NDARMA} show that for NDARMA models Cohen's $\kappa$
\[
\kappa(t) = \dfrac{\sum_{j=1}^r p_{jj}(t) - \sum_{j=1}^r p_j^2}{1 - \sum_{j=1}^r p_j^2}
\]is equivalent to $\text{corr}(X_1,X_{1+t})$ where $p_{jj}(t)$ is the probability state $j$ transitions to state $j$ in $t$ steps. An empirical (bias corrected) estimate of Cohen's $\kappa$ is provided by \cite{weiss:2008:NDARMA} \[
\hat{\kappa}(m) = 1 + \dfrac{1}{n} - \dfrac{1 - \sum_{j=1}^r \hat{p}_{jj}(m)}{1 - \sum_{j=1}^r\hat{p}_j^2}\label{chpt2eq5}
\] where $\hat{p}_{jj}(m)$ is the estimated proportion that state $j$ transitions to state $j$ in $m$ steps using all of the segments, i.e.\
\[
\hat{p}_{jj}(m) = \dfrac{1}{c}\sum_{i=1}^s \dfrac{1}{n_i-m}\sum_{t=2}^{n_i} Y^{(i)}_{tj}Y^{(i)}_{t+m,j}
\] The following proposition provides the asymptotic distribution for Pearson's chi-squared test of homogeneity corrected for autocorrelation induced in the data by the NDARMA model. 
%\cite{weiss:2009:cts} shows that the DAR(1) model is congruous to a Markov chain with transition matrix 
%\begin{equation}
%\begin{pmatrix}
%p_1(1-\phi) + \phi & p_1(1-\phi) & \hdots & p_1(1-\phi) \\
%p_2(1-\phi) & p_2(1-\phi) + \phi & & \vdots\\
%\vdots & & \ddots & p_{k-1}(1-\phi)\\
%p_k(1-\phi) & \hdots & p_k(1-\phi) & p_k(1-\phi) + \phi
%\end{pmatrix} \label{chpt2eq4}
%\end{equation}
%Wei\ss\ also shows that $\phi$ is the only non-unit eigenvalue of this transition matrix \cite[\S 11.2.9]{weiss:2009:cts}. \cite{weiss:2013:NDARMA} shows that a consistent and asymptotically normal estimator of $\phi$ is given by 
%\begin{equation}
%\hat{\phi} = 1 + \dfrac{1}{n} - \dfrac{1 - \sum_{j=1}^k \hat{p}_{jj}(1)}{1 - \sum_{j=1}^k\hat{p}_j^2}\label{chpt2eq5}
%\end{equation} 
%or a segment specific correlation parameter can be computed for each transition matrix
%\begin{equation}
%\hat{\phi}^{(i)} = 1 + \dfrac{1}{n} - \dfrac{1 - \sum_{j=1}^k \hat{p}^{(i)}_{jj}(1)}{1 - \sum_{j=1}^k(\hat{p}^{(i)}_j)^2}\label{chpt2eq6}
%\end{equation}
%where $\hat{p}^{(i)}_{jj}(m)$ is the estimated proportion that state $j$ transitions to state $j$ in $m$ steps for segment $i$ and 
%where $\hat{p}_{jj}(m)$ is the estimated proportion that state $j$ transitions to state $j$ in $m$ steps using all of the segments, i.e.\

%\[
%\hat{p}^{(i)}_{jj}(m) = \dfrac{1}{n-m}\sum_{t=2}^n Y^{(i)}_{tj}Y^{(i)}_{t+m,j} \qquad
%\hat{p}_{jj}(m) = \dfrac{1}{c}\sum_{i=1}^c\dfrac{1}{n_i-m}\sum_{t=2}^{n_i} Y^{(i)}_{tj}Y^{(i)}_{t+m,j}
%\]

\begin{proposition}[Test of Homogeneity]\label{prop1}
Let $X_t^{(i)}$ be a categorical time series which follows an NDARMA($p,q$) model with parameters $\mathbf{p}^{(i)} = (p_1^{(i)},\ldots,p_r^{(i)})^\intercal$ (unknown), $\bm{\phi} = (\phi_1,\ldots,\phi_p)$ (known), and $\bm{\varphi} = (\varphi_0,\ldots,\varphi_q)^\intercal$ (known) for $i=1,\ldots,s$. Then under the null hypothesis that $\mathbf{p} = \mathbf{p}^{(1)} = \cdots = \mathbf{p}^{(s)}$, $X^2/c$, where $X^2$ is Pearson's chi-squared test statistic, is asymptotically $\chi^2$ with degrees of freedom $(|R| -1)(s-1)$ and $c$ is given by Equation \ref{chpt2eqc}.
\end{proposition}

Proof of proposition \ref{prop1} is provided in the supplementary material. Note this proposition assumes the value of $c$ is known. In practice this is often unreasonable and an estimate of $c$ will have to be utilized. To assess convergence in MCMC output, assume the data come from a DAR(1) model, which is a subset of the NDARMA($p,q$) models. A categorical time series $X_t$ follows the DAR(1) model if the following recursion holds
\begin{equation}
X_t = \alpha_t X_{t-1} + \beta_t \epsilon_{t-1}
\label{chpt2eq3}
\end{equation}
where $(\alpha_t, \beta_t) \sim \text{Multinomial}(1, (\phi, 1-\phi))$ and $\epsilon_t \sim \text{Categorical}(p_1,\ldots,p_r)$ are independent for $t=1,\ldots,n$. The model implies that with some probability $\phi$, the current state of the categorical process is equal to the previous state, and with probability $1-\phi$ the current state is a draw from the marginal categorical distribution. The DAR(1) model has as its parameters the frequency distribution of the discrete categories, $p_1,\ldots,p_r$ as well as an autocorrelation parameter $\phi$. \cite{weiss:2013:NDARMA} shows that a consistent and asymptotically normal estimate of $\phi$ for a DAR(1) model is $\hat{\phi} = \hat{\kappa}(1)$. Additionally, \cite{weiss:2013:NDARMA}, shows that for the DAR(1) model the value $c$ reduces to \[
c = \dfrac{1+\phi}{1-\phi}
\] Therefore the Wei\ss\ procedure to assess convergence in MCMC output is to compute $X^2/\hat{c}$ and evaluate a $p$-value from the $\chi^2$ distribution with $(|R| -1)(s-1)$ degrees of freedom.

%Since the Tavar\'{e} result provides an asymptotic distribution that is not of closed form, simulation is required to obtain $p$-values. Evaluating this $p$-value is on the order of $O(B + k^c + n)$ where $B$ is the number of iterations to obtain the simulated $p$-value. We call the procedure where the $p$-value is evaluated using the estimate of the eigenvalues from the transition matrix the Tavar\'{e} procedure. We call the procedure using the DAR(1) model to estimate the eigenvalues the Wei\ss\ procedure. 

%Both the Tavar\'{e} and Wei\ss\ procedure can be computationally prohibitive especially when the number of categories or the number of chains is large. Two computationally efficient parametric bootstrap procedure for estimating the $p$-value are considered. At the end of section \ref{sec:meth} an overview of how to compute these procedures is provided. 

\subsubsection{Bootstrap Procedures}
When the assumptions of proposition \ref{prop1} are suspect, evaluating the $p$-value for the test statistic from a $\chi^2$ distribution may not be appropriate. In such situations performing one of the following bootstrap methods may be preferable. Both are parametric bootstrap methods. The first bootstrap method (DARBOOT) assumes the data arise from a DAR(1) model. The parameters of the DAR(1) model are estimated, $B$ bootstrap data replicates are generated using the estimated parameters substituted into Equation \ref{chpt2eq3}, and the corrected chi-square test statistic is evaluated for each generated data set. The second bootstrap method (MCBOOT) assumes the data arise from Markov chains of order 1, the transition matrix is estimated, $B$ bootstrap data replicates are generated using the estimated transition matrix, and the chi-square test statistic (Equation \ref{chpt2eq1}) is evaluated for each generated data set. For both procedures the bootstrap $p$-value is then estimated as the proportion of test statistics that are equal to or exceed the observed test statistic.

\subsection{Method 2: Transition Matrix}
\label{sec:met2}
Whereas Method 1 focuses on comparing the frequency distribution of discrete categories, Method 2 focuses on comparing transition probability matrices. Let the observed number of transitions from category $j$ to category $k$ in segment $i$ be $f_{jk}^{(i)}$. Then the test statistic is given by 
\begin{equation}
X^2_f = \sum_{i=1}^s\sum_{j=1}^r\sum_{k\in R_j} \dfrac{f_{j}^{(i)}\left(\hat{p}_{jk}^{(i)} - \hat{p}_{jk}\right)^2}{\hat{p}_{jk}}
\label{chpt2eq7}
\end{equation}
where $f_j^{(i)} = \sum_{k=1}^r f_{jk}^{(i)}$ is the total number of transitions from category $j$ in segment $i$, $\hat{p}_{jk} = \sum_{i=1}^s f_{jk}^{(i)} / \sum_{i=1}^s f_{j}^{(i)}$ is the pooled estimate of the transition probability from category $j$ to $k$, and $R_j = \left\{j|\hat{p}_{jk} > 0\right\}$ is the set of categories of nonzero observed transitions from category $j$. This is shown in \cite{billingsley:1961:MC} to have a $\chi^2$ distribution with $\sum_{j=1}^r (a_j-1)(b_j-1)$ degrees of freedom where $a_j$ is the number of unique transitions from state $j$, i.e. $a_j = |A_j|$ where $A_j = \left\{i:f_j^{(i)} > 0\right\}$ and $b_j$ is the number of positive entries in the $j$th row of the matrix for the entire sample, $b_j = |B_j|, B_j = \left\{k: \hat{p}_{jk} > 0\right\}$. 

Similar to the MCBOOT procedure, a bootstrap version of the Billingsley procedure may be carried out (BillingsleyBOOT). The Hangartner, Wei\ss, DARBOOT, MCBOOT, Billingsley, and BillingsleyBOOT procedures can be used to perform an $\alpha$-level test to determine whether the null hypothesis that the segments are from the same model can be rejected. Rejection of the null hypothesis is evidence that the chain has not converged to the target distribution. A summary of the specifics of evaluating these diagnostics is provided below:

\begin{itemize}
\item Method 1: Frequency Distribution
\begin{enumerate}[(1)]
\item Hangartner procedure
\begin{enumerate}[(i)]
\item Estimate the chain-specific probabilities $\hat{p}_j^{(i)}$ and pooled estimates $\hat{p}_j$.
\item Compute the test statistic $X^2$ from Equation \ref{chpt2eq1} and compute the $p$-value from a $\chi^2$ random variable with $(|R| -1)(s-1)$ degrees of freedom.
\end{enumerate}
\item Wei\ss\ procedure
\begin{enumerate}[(i)]
\item Obtain estimates of the parameters of the DAR(1) model: $\hat{p}_j^{(i)}$ and $\hat{\phi}$ for $i=1,\ldots,s$ and $j=1,\ldots,r$ using Equation \ref{chpt2eq3}. 
\item Compute the test statistic $X^2/\hat{c}$ and compute the $p$-value from a $\chi^2$ random variable with $(|R|-1)(s-1)$ degrees of freedom.
\end{enumerate}
\item DARBOOT procedure
\begin{enumerate}[(i)]
\item Obtain estimates of the parameters of the DAR(1) model: $\hat{p}_j^{(i)}$ and $\hat{\phi}$ for $i=1,\ldots,s$ and $j=1,\ldots,r$ using Equation \ref{chpt2eq3}. 
\item Simulate $B$ sets of parallel MCMC chains of the same number and length as the original chains using Equation \ref{chpt2eq3}.
\item Compute the test statistic $X^2$ from equation \ref{chpt2eq1} for each of the $B$ bootstrap samples, say $X^2_b$ for $b=1,\ldots,B$, and compute the $p$-value \[
p = \dfrac{1}{B}\sum_{b=1}^B \mathbbm{1}\left\{ X^2_b \geq X^2\right\}.
\]
\end{enumerate}
\item MCBOOT procedure
\begin{enumerate}[(i)]
\item Estimate the transition matrix for each chain using $\hat{p}_{jk}^{(i)} = f_{jk}^{(i)} / f_{j}^{(i)}$ where \[
f_{jk}^{(i)} = \sum_{t=2}^n Y_{tj}^{(i)}Y_{tk}^{(i)} \qquad f_{j}^{(i)} = \sum_{k=1}^r f_{jk}^{(i)}.
\]
\item Simulate $B$ sets of parallel MCMC chains of the same number and length as the original chains using the estimated transition matrix.
\item Compute the test statistic $X^2$ from equation \ref{chpt2eq1} for each of the $B$ bootstrap samples, say $X^2_b$ for $b=1,\ldots,B$, and compute the $p$-value as above in the DARBOOT procedure.
\end{enumerate}
\end{enumerate}
\item Method 2: Transition Matrix
\begin{enumerate}[(1)]
\item Billingsley procedure
\begin{enumerate}[(i)]
\item Estimate the transition matrix for each chain using $\hat{p}_{jk}^{(i)} = f_{jk}^{(i)} / f_{j}^{(i)}$.
\item Compute the test statistic $X^2_f$ from Equation \ref{chpt2eq7} and compute the $p$-value from a $\chi^2$ random variable with $\sum_{j=1}^r (a_j-1)(b_j-1)$ degrees of freedom.
\end{enumerate}
\item  BillingsleyBOOT
\begin{enumerate}[(i)]
\item Estimate the transition matrix for each chain using $\hat{p}_{jk}^{(i)} = f_{jk}^{(i)} / f_{j}^{(i)}$.
\item Simulate $B$ sets of parallel MCMC chains of the same number and length as the original chains using the estimated transition matrix.
\item Compute the test statistic $X^2_f$ from equation \ref{chpt2eq7} for each of the $B$ bootstrap samples, say $X^2_{fb}$ for $b=1,\ldots,B$, and compute the $p$-value above in the DARBOOT procedure.
\end{enumerate}
\end{enumerate}
\end{itemize}

\section{Simulation}
\label{sec:sim}
Simulation is used to assess the performance of the diagnostics. In the simulation, two independent segments of length $t$ are generated. The first segment is simulated from a DAR(1) model with parameters $\phi$ and marginal probability distribution $\mathbf{p}$. The second segment is simulated from a DAR(1) model with parameters $\phi$ and marginal probability distribution $\beta\mathbf{p} + (1-\beta)\mathbf{q}$. The segment length is varied as $t=10, 100, 1000, 10000$; the autocorrelation parameter $\phi$ is varied as $\phi = 0.0, 0.25, 0.5, 0.75$. The marginal probabilities are $\mathbf{p} = (0.25, 0.3, 0.45)^\intercal$ and $\mathbf{q} = (0.75, 0.05, 0.2)^\intercal$ with $\beta$ ranging in 
\[
\beta = 0.0,  0.3, 0.5, 0.7, 0.8, 0.85, 0.9, 0.94, 0.96, 1.00
\] When $\beta = 1.0$ the two segments are from the same model, and when $\beta=0.0$ they are from two completely distinct models. When $\beta \in (0,1)$ the second segment is a convex combination of these two models. A total of $N=1000$ simulations are run and Methods 1 and 2 computed at each iteration. 

\begin{figure}[ht!]
\centering
\includegraphics[width=0.90\textwidth,keepaspectratio]{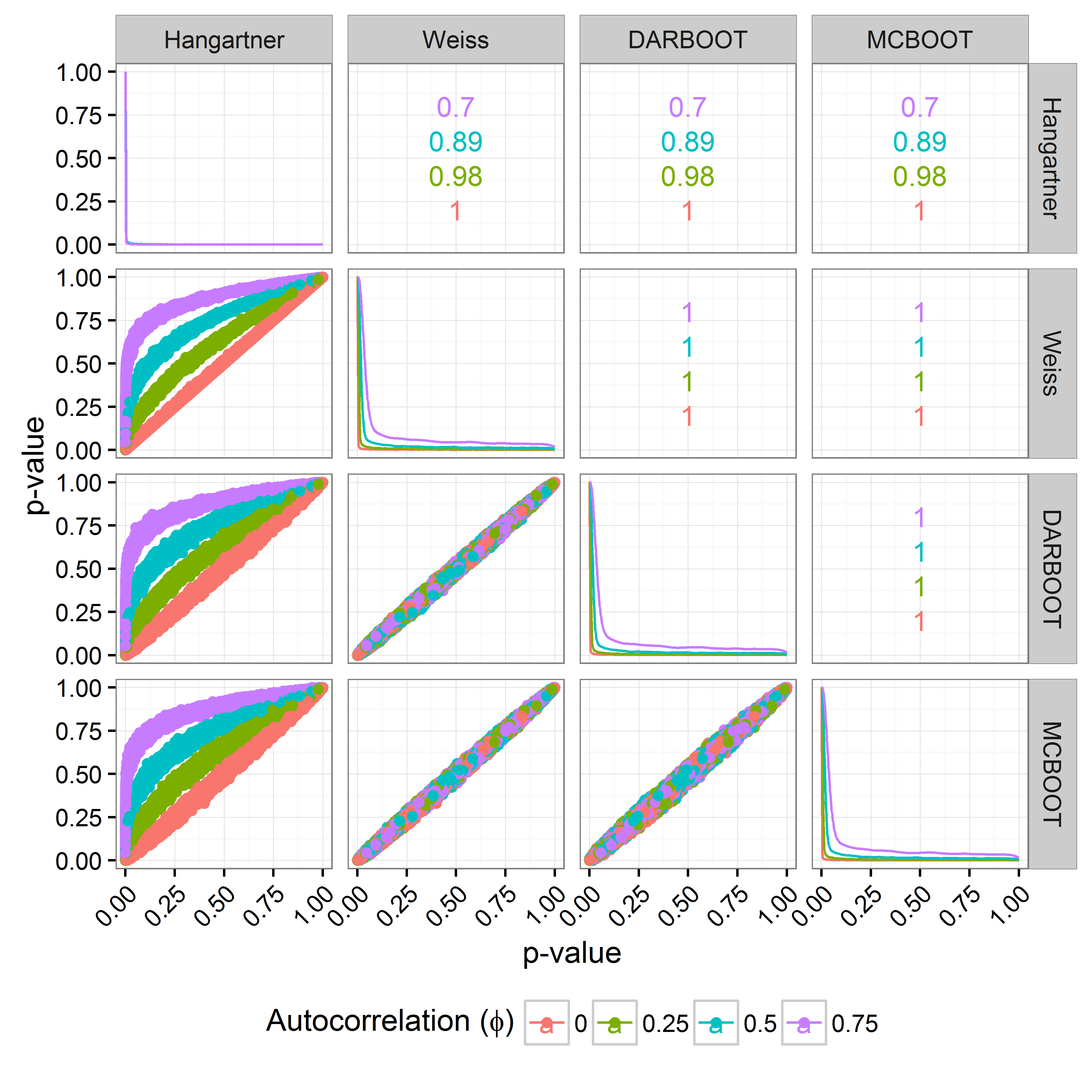}
\caption{The concordance of Method 1 procedures. Only segment lengths greater than 100 are plotted. Colors correspond to autocorrelation $\phi$ where red is $\phi=0$, green is $\phi=0.25$, blue is $\phi=0.5$, and purple is $\phi=0.75$.}
\label{f0}
\end{figure}

\begin{figure}[ht!]
\centering
\includegraphics[width=\textwidth,keepaspectratio]{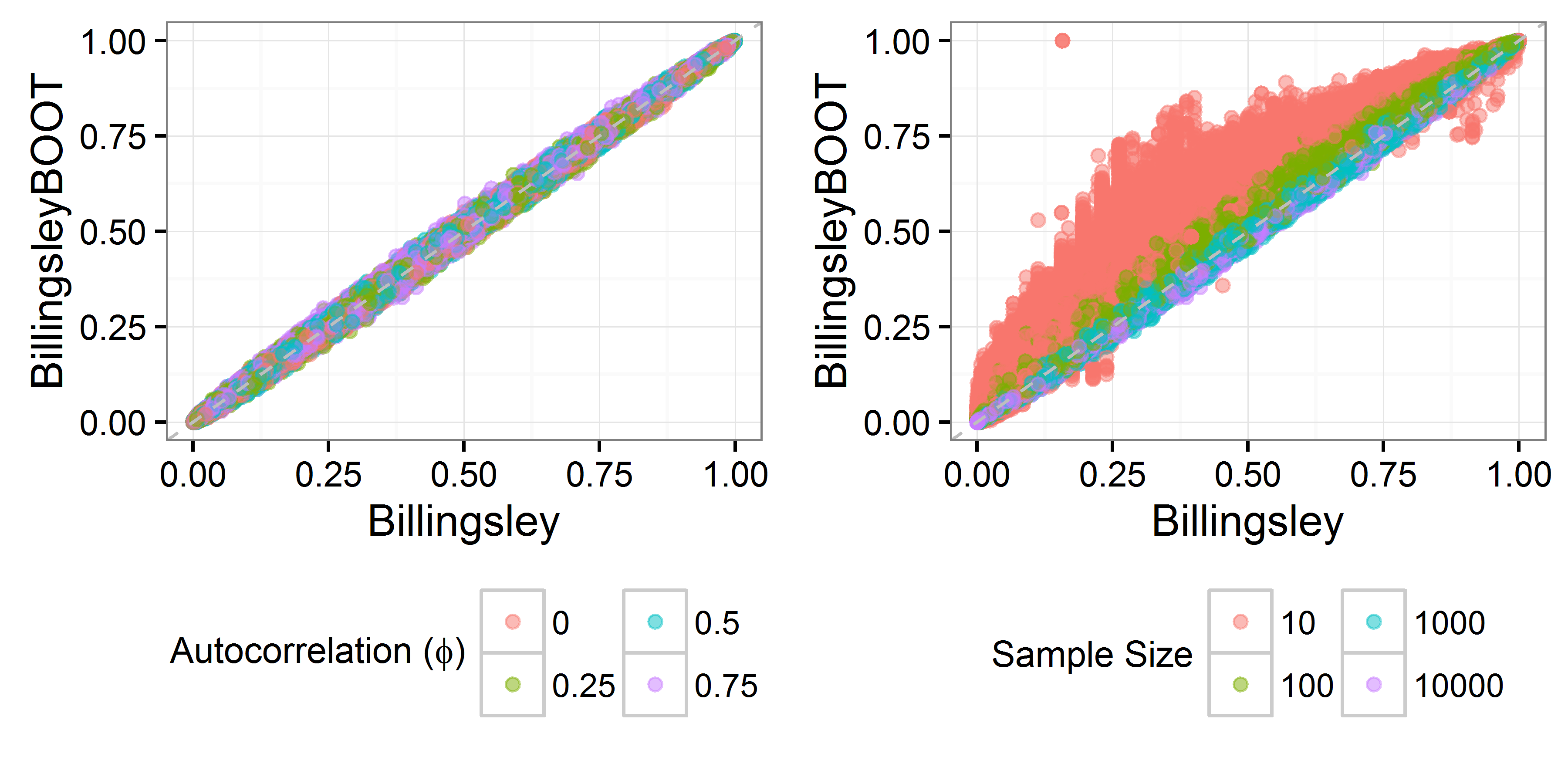}
\caption{The concordance of Method 2 procedures. Only segment lengths greater than 100 are plotted. Colors correspond to autocorrelation $\phi$ where red is $\phi=0$, green is $\phi=0.25$, blue is $\phi=0.5$, and purple is $\phi=0.75$.}
\label{f0b}
\end{figure}

\begin{figure}[ht!]
\centering
\includegraphics[width=\textwidth,keepaspectratio]{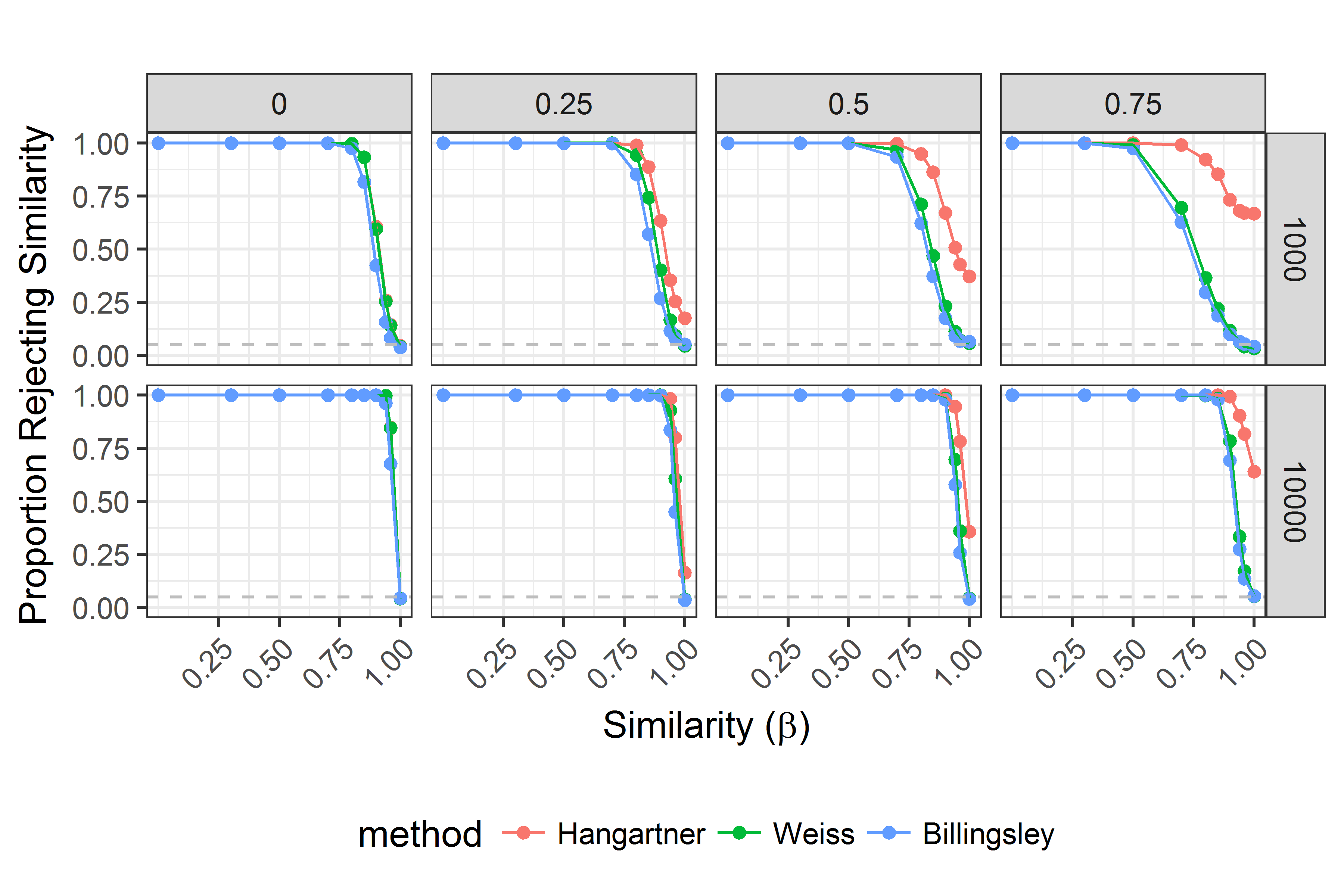}
\caption{Convergence diagnostics operating characteristics. The vertical axis represents the proportion of simulations for which the diagnostic did not reject. Horizontal axis is the similarity of the two segments, i.e.\ $\beta=1.0$ means they are from same model. The columns correspond to values of autocorrelation $\phi$ and the rows correspond to the segment length $t$.}
\label{f1}
\end{figure}

Below the main diagonal of the square matrix of plots in Figure \ref{f0}, the $p$-values of Method 1 procedures are plotted against each other. On the main diagonal of Figure \ref{f0}, the distribution of the $p$-values is plotted. The plots above the main diagonal display the correlation between the procedures. Figure \ref{f0} shows that all of the procedures except for the Hangartner procedure are highly correlated. The Hangartner procedure correlated well when there is no autocorrelation in the data ($\phi=0.0$). However when there is autocorrelation in the data the Hangartner procedure results in overestimated $p$-values and this bias increases as the autocorrelation increases. The asymptotic result from the Wei\ss\ method correlates well with the bootstrap methods. The non-Hangartner procedures all have correlation close to 1 with each other at all levels of autocorrelation. The Hangartner procedure's correlation with the other methods ranges from 0.7 at high autocorrelation to 1 at no autocorrelation. 

Figure \ref{f0b} shows the correlation of the $p$-values for procedures in Method 2. There is high correlation between the bootstrap method and the asymptotic result. The correlation is near 1 for all values of autocorrelation. 

Figure \ref{f1} displays the operating characteristics of the diagnostics on the simulated data. The vertical axis represents the proportion of simulations for which the diagnostic rejects the null hypothesis that the two segments are independent. The horizontal axis is the $\beta$ value. In Figure \ref{f1} when $\beta < 1$ the curves represent the power of the diagnostics to identify a difference in the two independent segments. When $\beta=1$, the curve represents the type I error, i.e.\ the probability a diagnostic incorrectly rejects the null.

When there is no autocorrelation in the model ($\phi=0$), all of the diagnostics perform well: the diagnostics have high power and type I error rate is around $\alpha = 0.05$. However, as the autocorrelation increases, the power of the tests to differentiate the two segments drops. At high levels of autocorrelation ($\phi = 0.75$) the two segments have to be quite different for the diagnostics to maintain high power. This effect is attenuated as the segment length increases. 

More important is the behavior of the diagnostics at $\beta=1$ as this is when the two segments are derived from the same model. When $\beta=1$ this simulates the situation where the MCMC algorithm has converged to the target distribution. The diagnostics should not reject the null hypothesis that the segments are similar. Figure \ref{f1} shows that all diagnostics have this behavior except for the Hangartner diagnostic, which does not take into account the autocorrelation. 

\section{Real data analysis}
\label{sec:data}
The enzymatic activity data set \citep[\S 4.1]{green:1997:nmix} is used to compare the diagnostics developed in Section \ref{sec:meth} to the method developed by \cite{sisson:2007:conv}. The enzymatic activity data are the distribution of enzymatic activity in the blood for an enzyme involved in metabolism of carcinogenic substances among a group of 245 unrelated individuals. In \cite{green:1997:nmix} a finite component normal mixture model is fit to the data and MCMC used to obtain samples from the posterior of the parameters involved. In brief the model is given by 
\[
y_i \sim \sum_{j=1}^k w_j f(\cdot|\theta_j) \qquad\text{independently for $i=1,\ldots,n$}
\] where $k$ is an unknown number of mixture components and $f(\cdot|\theta)$ is a given parametric family of densities indexed by parameter $\theta$. Green uses the Normal distribution with $\theta_j = (\mu_j, \sigma^2_j)$. In the finite mixture model $k$ is the discrete parameter. Samples are drawn from the posterior of the finite mixture model parameters using Green's RJMCMC sampler. Five independent chains are produced with five million iterations. No burn-in or thinning was used. Sampling was done by software provided by \cite{green:1997:nmix}. 

\begin{figure}[t!]
\centering
\includegraphics[width=\textwidth,keepaspectratio]{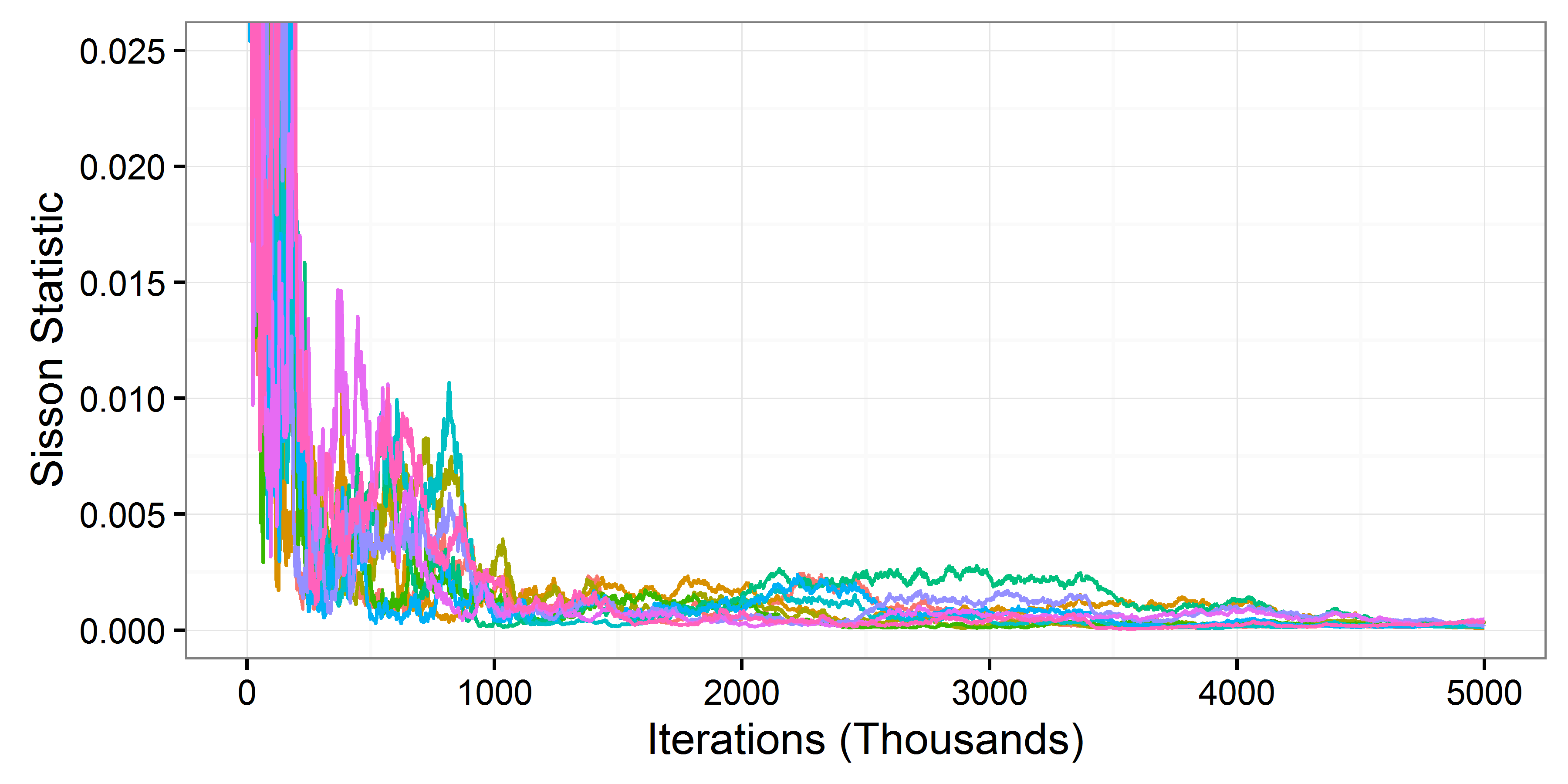}
\caption{Sisson and Fan diagnostic for MCMC chains from the Bayesian model of Richardson and Green fit to the enzymatic activity data. Each line measures the discrepancy of one chain to another (the closer to 0 the more similar the chains are).}
\label{f3}
\end{figure}

The output of the diagnostic of \cite{sisson:2007:conv} is presented in Figure \ref{f3}. See Section \ref{sec:other} for a description of this method. The method by \cite{sisson:2007:conv} does not directly compare the discrete variable $k$ between chains, but rather uses a surrogate measure by comparing the distance of the continuous variables to predefined reference points. Each line in Figure \ref{f3} measures the discrepancy of one chain to another, and the closer to 0 the more similar the chains are. This plot shows there is substantial variation between the chains up to a million iterations. Beyond a million iterations a few chains seem to diverge a bit, but overall the diagnostic plot suggests that the MCMC algorithm has converged to the target distribution. 

\begin{figure}[t!]
  \centering
\includegraphics[width=0.9\textwidth,keepaspectratio]{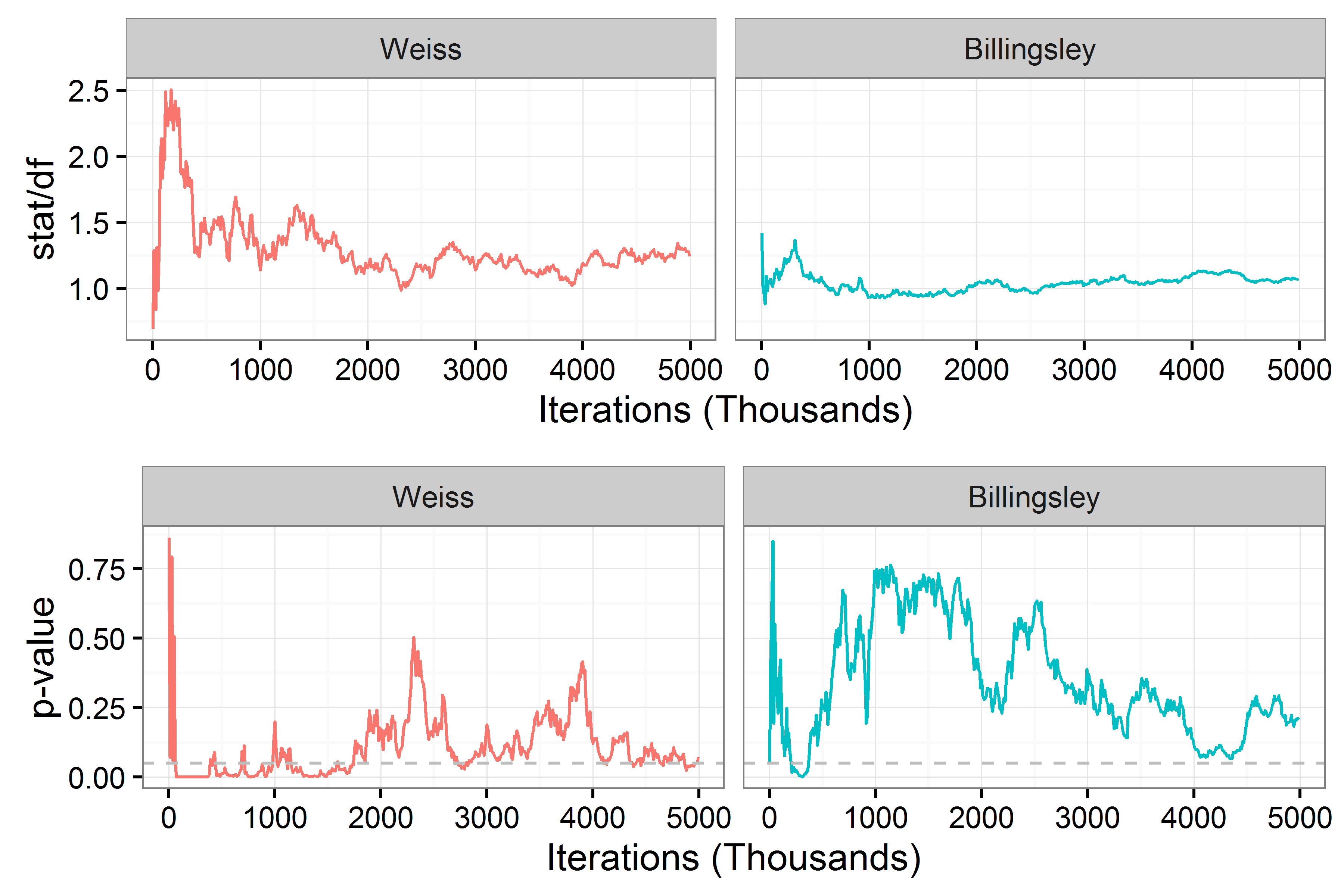}
\caption{MCMC convergence assessment for  the discrete model parameter in the enzymatic activity data analysis. Horizontal axis is the MCMC iterations. In the first row the vertical axis is the test statistic. In the second row the vertical axis is the $p$-value.}
\label{m1}
\end{figure}

The methods from Section \ref{sec:meth} are applied to this dataset, using Weiss procedure for Method 1 and the asymptotic Billingsley procedure for Method 2 (see Section \ref{sec:disc} for a discussion on which procedures are recommended). The results are in Figure \ref{m1}. The first row represents the test statistic. The second row is the $p$-value associated with these test statistics. In the second row the value $p=0.05$ is indicated by a gray dashed line. The first column is Method 1, and the second column is Method 2. Both Method 1 and Method 2 show very high values for the test statistics before a million iterations, similar to the results from Fan and Sisson. This example provides evidence that the methods developed in Section \ref{sec:meth} are in agreement with those developed by \cite{sisson:2007:conv}. 

\section{Software}

Software to evaluate these convergence diagnostics is available in the Mamba package, a package for Bayesian analysis in the julia language developed by \cite{smith:2014:Mamba}. One function is available to evaluate the diagnostics presented above with a keyword argument to specify which procedure to use. For Method 1 the user can select from the Hangartner, Weiss, DARBOOT, and MCBOOT procedures with Weiss selected as default. For Method 2 the user can select from the Billingsley and BillingsleyBOOT procedure with the Billingsley procedure as default. For both methods results for within chain and between chain convergence are reported. Users can select the portion of the tails of the chains that are used for the within chain diagnostic (default is 30\%). An option to plot the between chain diagnostic (as in Figure \ref{m1}) is also available. 

\begin{figure}
  \centering
\includegraphics[width=0.8\textwidth,keepaspectratio]{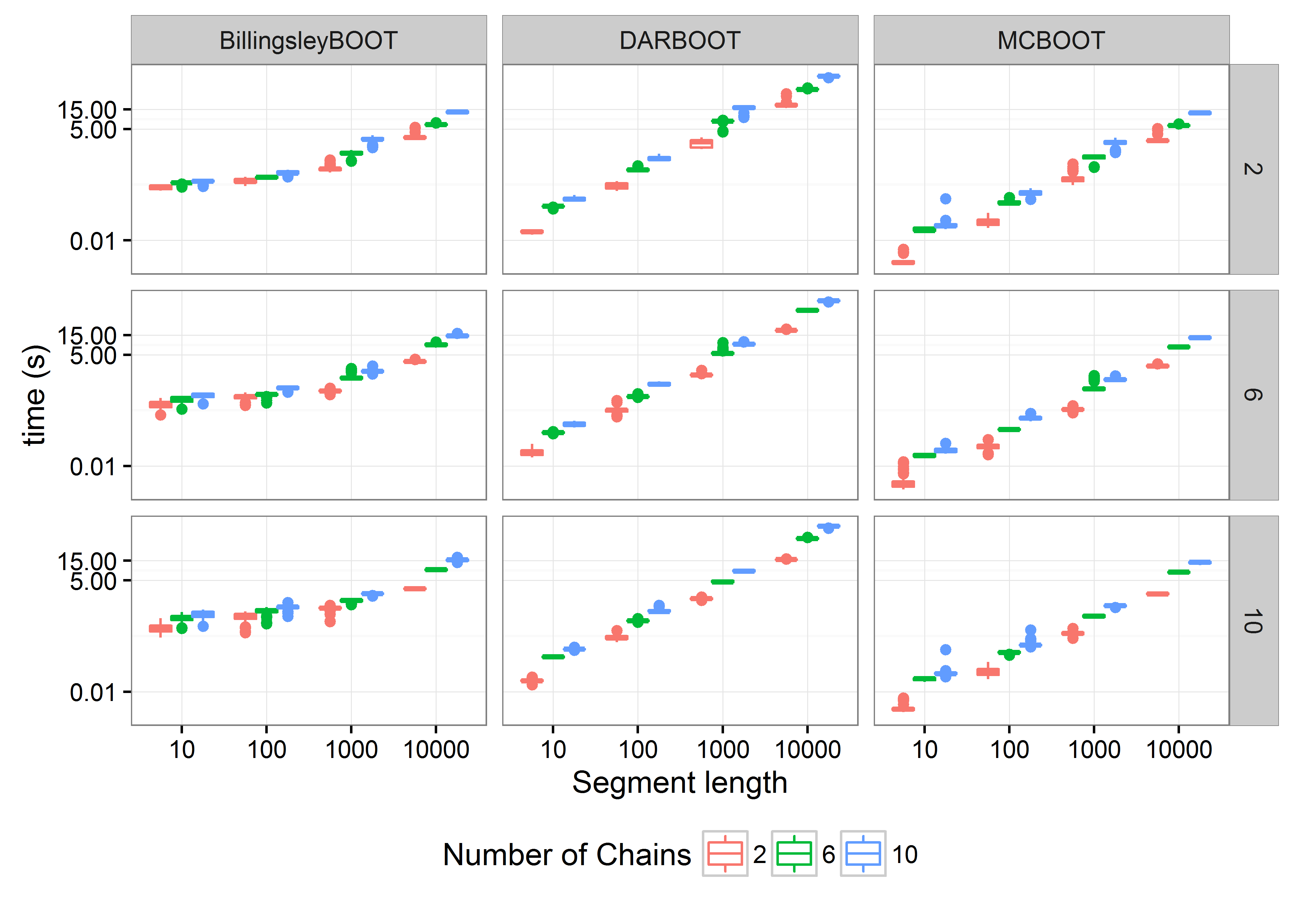}
\caption{Timing results on simulated data for the bootstrapped diagnostics. Horizontal axis denotes segment length, vertical axis is time, rows correspond to number of categories, columns to procedure, and colors to number of chains.}
\label{f81}
\end{figure}

\begin{figure}
  \centering
\includegraphics[width=0.8\textwidth,keepaspectratio]{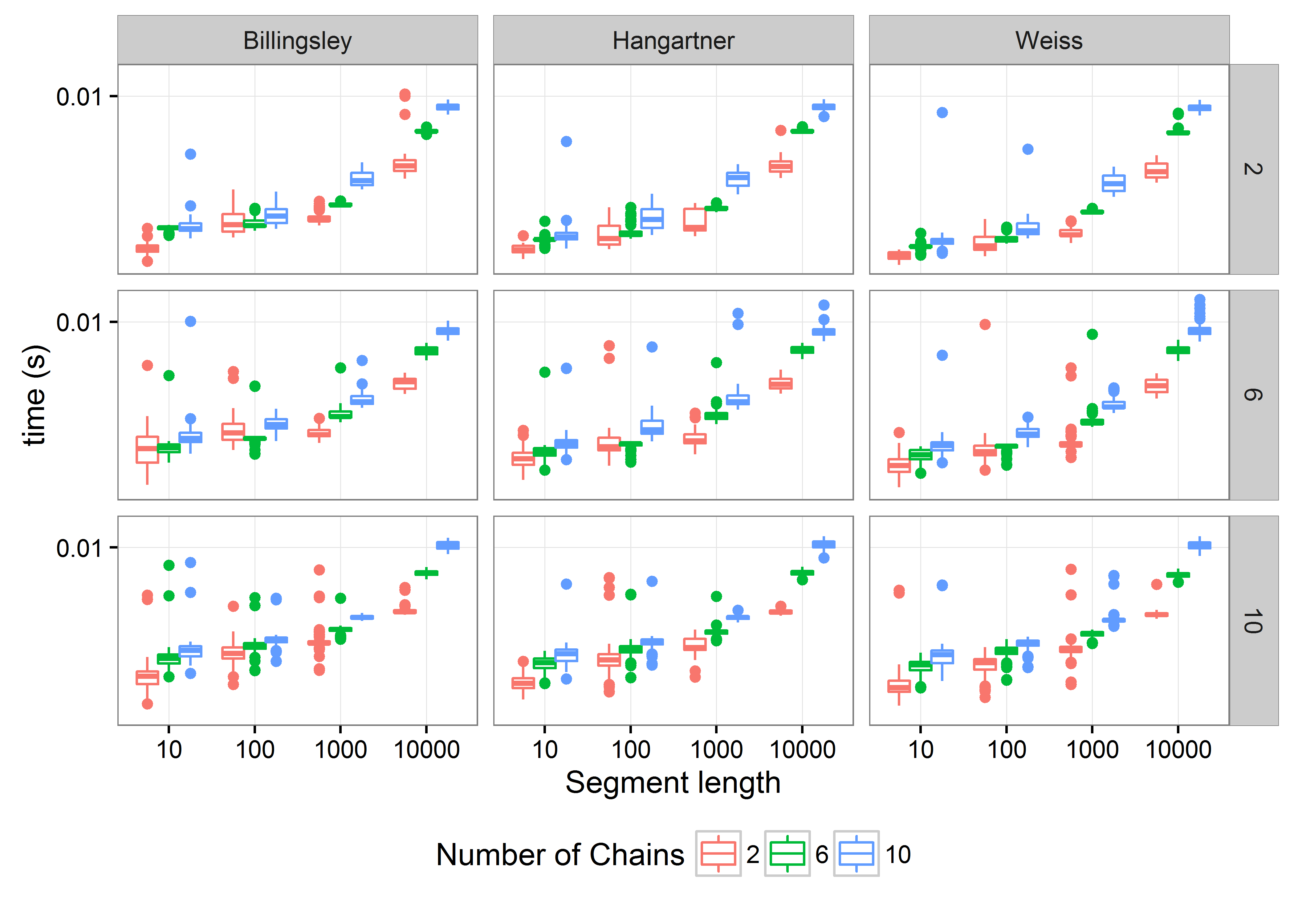}
\caption{Timing results on simulated data for the asymptotic diagnostics. Horizontal axis denotes segment length, vertical axis is time, rows correspond to number of categories, columns to procedure, and colors to number of chains.}
\label{f82}
\end{figure}

Timing results for the function are presented in Figure \ref{f81} and Figure \ref{f82} (supplementary information). The plots display boxplots of timing results for the discrete diagnostics. For the simulation $c$ chains were simulated from a DAR(1) model where $c$ ranged from two to ten (by two), the number of categories $k$ in the DAR(1) model ranged from from two to ten (by two), and the segment length of the chains was one of 10, 100, 1000, 10000. At each combination of variables one hundred simulations were performed. 

Figure \ref{f81} features the bootstrapped procedures and Figure \ref{f82} features the asymptotic procedures.  The figure shows that the asymptotic procedures (Hangartner, Weiss, and Billingsley) are all comparable in their run time. The ordering of the bootstrap procedures is BillingsleyBOOT (slowest), followed by DARBOOT and MCBOOT. The asymptotic procedures were significantly faster than the bootstrap procedures. There is a clear linear increase in the runtime as the number of chains increases. Increasing the number of categories has a much smaller impact on runtime.

\section{Discussion}
\label{sec:disc}

Assessing whether an MCMC procedure has converged to the target distribution is important because performing any kind of inference assumes the MCMC output are draws from the target distribution. Classic convergence diagnostics are non-optimal for categorical data because they depend on estimation of spectral density, on suspect normality assumptions, or on determining overdispersion within a small number of outcomes. The methods presented in this paper are built for discrete data from MCMC output and make little assumptions about the structure of the data. The only necessary assumption is that the data come from reversible Markov chains, which holds for most MCMC algorithms. Simulation results indicate that ignoring the dependence in MCMC output is not appropriate. Finally, applying these methods to MCMC output from Green's reversible jump MCMC sampler provides comparable results to a convergence diagnostic tailor made for that sampler. 

Due to the discordance between the Hangartner procedure and the other Method 1 procedures, the Hangartner procedure is not recommended. Due to its faster computational speed the Weiss procedure is recommended for Method 1. For Method 2, the asymptotic Billingsley approach is recommended because it has high agreement with the more computationally expensive bootstrap method. 

\begin{figure}
\centering
\includegraphics[width=0.8\textwidth,keepaspectratio]{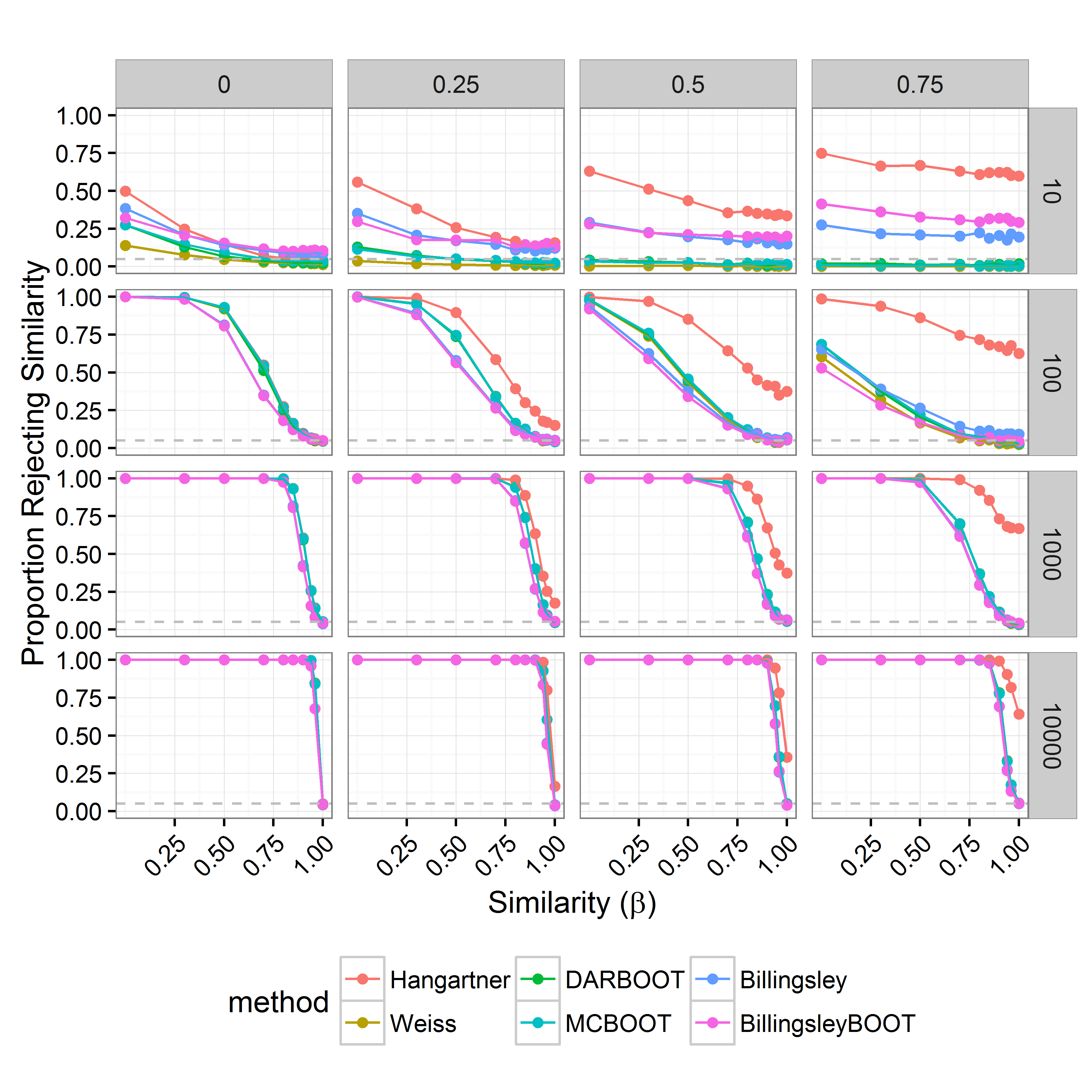}
\caption{Convergence diagnostics operating characteristics. The vertical axis represents the proportion of simulations for which the diagnostic did not reject. Horizontal axis is the similarity of the two segments, i.e.\ $\beta=1.0$ means they are from same model. The columns correspond to values of autocorrelation $\phi$ and the rows correspond to the segment length $t$.}
\label{f1s}
\end{figure}

Figure \ref{f1s} (supplementary information) demonstrates that the diagnostic methods do not perform well when segment length is less than or equal to 100 (low power). This is typically not an issue since MCMC output is often much longer. Nonetheless these methods are not recommended for short runs of MCMC output. It is worth noting however that the error made by these methods even with low segment length is conservative. That is, if these test statistics were used to assess convergence diagnostics, and the MCMC output was not very long, the methods will suggest that more iterations need to be obtained (even if MCMC output has converged to sampling from the target distribution). This is a safer error than assessing convergence when the output has not actually converged. 

This paper described several procedures for assessing the convergence of MCMC output for discrete parameters. The lack of convergence diagnostics for discrete parameters, during a time in which models with discrete parameters are quite popular, reveals the timeliness of the methods presented. Models including discrete parameters will be greatly benefited by the additional convergence checks.

\section{SUPPLEMENTARY MATERIAL}

\subsection{NDARMA Model}
The NDARMA model was first described by \cite{jacobs:1983:NDARMA}. The following definition of the NDARMA model is given by \cite{weiss:2008:NDARMA} in which it was also proved to be congruous with Jacobs' original definition. 

Let $\left\{X_t\right\}_{\mathbb{Z}}$ and $\left\{\epsilon_t\right\}_{\mathbb{Z}}$ be categorical processes with support $\mathcal{V}=\left\{1,\ldots,r\right\}$. Let $\left\{\epsilon_t\right\}_{\mathbb{Z}}$ be independent and identically distributed (i.i.d.) with marginal distribution \[
\text{Categorical}(p_1,\ldots,p_r)
\] Each $\epsilon_t$ is assumed independent of $\left\{X_s\right\}_{s< t}$.

Define the i.i.d.\ random vectors $\mathbf{D}_t = (\alpha_{t,1},\ldots,\alpha_{t,p},\beta_{t,0},\ldots,\beta_{t,q})$ $$\mathbf{D}_t \sim \text{Multinomial}(1, \phi_1,\ldots,\phi_p,\varphi_0,\ldots,\varphi_q)$$ for $t\in \mathbb{Z}$, $\varphi_q > 0$ and $\varphi_0 > 0$ if $p\geq 1$. Each $\mathbf{D}_t$ is independent of $\left\{\epsilon_t\right\}_{\mathbb{Z}}$ and $\left\{X_s\right\}_{s<t}$. The process $\left\{X_t\right\}_{\mathbb{Z}}$ is said to be an NDARMA($p,q$) process if it follows the recursion $$X_t = \sum_{i=1}^p \alpha_{t,i} X_{t-i} + \sum_{j=0}^q \beta_{t,j} \epsilon_{t-j}$$ In the case of $q=0$, the process is said to be a DAR($p$) process. In the case of $p=0$ it is said to be a DMA($q$) process. 

\subsection{Proof of proposition \ref{prop1}}
The proof of the asymptotic distribution of the Pearson chi-squared test of homogeneity, under the assumption that data arise from an NDARMA process, follows from the proof of the classical result for independent data by Cram\'{e}r along with additional results by Jacobs and Lewis. Page 426-434 of \cite{cramer:1946:mms} provides a proof of the asymptotic distribution of the Pearson chi-squared statistic for goodness of fit with estimated parameters. The proof which follows relies on the generalization of proof for goodness of fit to the test of homogeneity found on page 446 of \cite{cramer:1946:mms}. 

First the Test of Homogeneity result using independent data is stated 

\begin{proposition}[Test of Homogeneity] 
For $i=1,\ldots,s$, let $X_t^{(i)}$ be a categorical sequence of length $n_i$ (indexed by $t$) that takes on values in $\mathcal{V} = \left\{1,\ldots,r\right\}$ such that $\text{Pr}\left\{ X_t^{(i)} = j\right\} = p_j^{(i)}$. Assume that $p_j^{(i)} = p_j$ for $j=1,\ldots,r-1$, $p_r^{(i)} = 1 - \sum_{j=1}^{r-1} p_j$, $p_j^{(i)}$ has continuous first and second derivatives with respect to the $p_j$, and that the matrix of first derivatives $\partial p_j^{(i)} / \partial p_j$ is of rank $r-1$. Then the system of equations \[
 \sum_{i=1}^s \sum_{j=1}^{r} \dfrac{N_j^{(i)} - n_ip_j}{p_j}\dfrac{\partial p_j^{(i)}}{\partial p_k}
\] for $k=1,\ldots,r-1$, referred to as the modified $\chi^2$ minimum equations, has one solution $\hat{\mathbf{p}}$ that converges in probability to the true $\mathbf{p}$ where $\hat{\mathbf{p}} = (\hat{p}_1,\ldots,\hat{p}_{r-1})^\intercal$ and $\mathbf{p} = (p_1,\ldots,p_{r-1})^\intercal$. The Pearson chi-square test statistic with this estimate of $\mathbf{p}$ \[
X^2 = \sum_{i=1}^s \sum_{j=1}^r \dfrac{n_i(\hat{p}_j^{(i)} - \hat{p}_j)^2}{\hat{p}_j}
\] is asymptotically distributed as $\chi^2$ random variable with $rs-(r-1)-s = (r-1)(s-1)$ degrees of freedom. 
\end{proposition}

Cram\'{e}r outlines the proof for this result \citep[pg. 445]{cramer:1946:mms}. \cite{jacobs:1978:dts2} extends Cram\'{e}r's goodness of fit result to handle data that follow the DARMA(1,q) model (a subset of the more general NDARMA models). \cite{weiss:2008:NDARMA} extends the result to the NDARMA model. This proof extends Cram\'{e}r's result to a test of homogeneity in data that follow the NDARMA model.
Let $X_t^{(i)}$ be categorical time series of length $n_i$ which follow an NDARMA($p,q$) model with parameters $\mathbf{p}^{(i)} = (p_1^{(i)},\ldots,p_r^{(i)})^\intercal$ (unknown), $\bm{\phi} = (\phi_1,\ldots,\phi_p)^\intercal$ (known), and $\bm{\varphi} = (\varphi_0,\ldots,\varphi_q)^\intercal$ (known) for $i=1,\ldots,s$. Under the null hypothesis that each categorical time series comes from the same NDAMRA model, the $p_j^{(i)}$ can be parametrized by the following $r-1$ constants 
\[
p_j^{(i)} = p_j
\] for $j=1,\ldots,r-1$ and $i=1,\ldots,s$. Let $p_r^{(i)} = p_r = 1 - \sum_{j=1}^{r-1} p_j$ then the partial derivatives are 
\begin{equation}
\dfrac{\partial p_j^{(i)}}{\partial p_k} = \begin{cases}
1 &\text{if $j=1,\ldots, r-1$ and $j=k$}\\
0 &\text{if $j=1,\ldots, r-1$ and $j\neq k$}\\
-1&\text{if $j=r$}
\end{cases} \label{peq1}
\end{equation} using notation from Section \ref{sec:meth} the modified $\chi^2$ equations become \begin{align*}
  \sum_{i=1}^s \sum_{j=1}^{r} \dfrac{N_j^{(i)} - n_ip_j}{p_j}\dfrac{\partial p_j^{(i)}}{\partial p_k} &= 0 &\text{for $k=1,\ldots,r$}\intertext{Using Equation \ref{peq1} this reduces to}
\sum_{i=1}^s \dfrac{N_j^{(i)} - n_ip_j}{p_j} &= 0 &\text{for $j=1,\ldots,r$}\\
\sum_{i=1}^s \dfrac{N_j^{(i)}}{p_j} &= \sum_{i=1}^s n_i\\
p_j &= \sum_{i=1}^s N_j^{(i)} \big/ \sum_{i=1}^s n_i
\end{align*} which is the estimator $\hat{p}_j$ used in Section \ref{sec:meth}. Define \[
x_{ij} = \dfrac{N_j^{(i)} - n_ip_j}{\sqrt{n_ip_j}} \qquad y_{ij} = \dfrac{N_j^{(i)} - n_i\hat{p}_j}{\sqrt{n_i\hat{p}_j}}
\] and $rs\times 1$ vectors \[
\mathbf{x} = \begin{pmatrix} \mathbf{x}_1 \\ \vdots \\ \mathbf{x}_s\end{pmatrix} \quad \mathbf{y} = \begin{pmatrix} \mathbf{y}_1 \\ \vdots \\ \mathbf{y}_s\end{pmatrix}
\] where $\mathbf{x}_i = (x_{i1},\ldots,x_{ir})^\intercal$ and $\mathbf{y}_i = (y_{i1},\ldots,y_{ir})^\intercal$. Let $\mathbf{B}$ be a $rs \times r-1$ matrix with elements equal to $p_j^{-1/2}\partial p_{j}^{(i)}/\partial p_k$. So $\mathbf{B}$ is a block matrix where the blocks are vertically stacked \begin{align*}
\mathbf{B} &= \begin{pmatrix} \mathbf{B}_1 \\ \vdots \\ \mathbf{B}_s \end{pmatrix}\intertext{where the $i$th block for $i=1,\ldots,s$ is}
\mathbf{B}_i &= \begin{pmatrix}
\dfrac{1}{\sqrt{p_1}}\dfrac{\partial p_{1}^{(i)}}{\partial p_1} & \hdots & \dfrac{1}{\sqrt{p_{1}}}\dfrac{\partial p_{1}^{(i)}}{\partial p_{r-1}}\\
\vdots & \ddots & \vdots\\
\dfrac{1}{\sqrt{p_{r-1}}}\dfrac{\partial p_{r-1}^{(i)}}{\partial p_1} & \hdots & \dfrac{1}{\sqrt{p_{r-1}}}\dfrac{\partial p_{r-1}^{(i)}}{\partial p_{r-1}}\\
\dfrac{1}{\sqrt{1-\sum_{j=1}^{r-1}p_j}}\dfrac{\partial p_{r}^{(i)}}{\partial p_1} & \hdots & \dfrac{1}{\sqrt{1-\sum_{j=1}^{r-1} p_j}}\dfrac{\partial p_{r}^{(i)}}{\partial p_{r-1}}
\end{pmatrix}\\
&= 
\begin{pmatrix}
    \dfrac{1}{\sqrt{p_1}} & \multicolumn{2}{c}
          {\text{\kern0.5em\smash{\raisebox{-1ex}{\Large 0}}}} \\
        & \ddots &  \\
      \multicolumn{2}{c}{\text{\kern-0.5em\smash 
      {\raisebox{0.75ex}{\Large 0}}}} & \dfrac{1}{\sqrt{p_{r-1}}}\\
\dfrac{-1}{\sqrt{1-\sum_{j=1}^{r-1}p_j}} & \hdots & \dfrac{-1}{\sqrt{1-\sum_{j=1}^{r-1} p_j}}
\end{pmatrix}
\end{align*} Since $\hat{p}_1,\ldots,\hat{p}_{r-1}$ is the solution to the modified $\chi^2$ equations, the second part of Cram\'{e}r's proof shows that $\mathbf{y} = \mathbf{A}\mathbf{x} + \mathbf{e}$ where $\mathbf{A} = \mathbf{I}_{rs} - \mathbf{B}(\mathbf{B}^\intercal\mathbf{B})^{-1}\mathbf{B}^\intercal$ and $\mathbf{e}$ tends in probability to zero ($\mathbf{I}_k$ is the identity matrix of dimension $k\times k$). 

Next we show that $\mathbf{x}$ is asymptotically normal and hence $\mathbf{y}$ is likewise asymptotically normal. Let $Z_{tj}^{(i)} = Y_{tj}^{(i)} - p_j$ and $\mathbf{Z}^{(i)} = (Z_{t1}^{(i)}, \ldots, Z_{tr}^{(i)})^\intercal$. Note that 
\begin{align*}
\text{E}[Z^{(i)}_{1,j}Z^{(i)}_{1+t,k}] &= \text{E}[(Y^{(i)}_{1,j} - p_j)(Y^{(i)}_{1+t,k} - p_k)]\\
&= \text{E}[Y^{(i)}_{1,j}Y^{(i)}_{1+t,k}] - p_j\text{E}[Y^{(i)}_{1+t,k}] - p_k\text{E}[Y^{(i)}_{1,j}] + p_jp_k\\
&= p_{jk}(t) - p_jp_k - p_jp_k + p_jp_k\\
&= p_{jk}(t) - p_jp_k \intertext{\cite[pg. 229]{weiss:2013:NDARMA} shows this is equivalent to}
&=p_j(\delta_{jk} - p_k)\text{corr}(X^{(i)}_1,X^{(i)}_{1+t})
\end{align*} where $\delta_{jk}$ is 1 if $j=k$ and 0 otherwise. \cite{weiss:2013:NDARMA} shows that $\mathbf{Z}^{(i)}$ is stationary, $\alpha$-mixing, and with $\text{E}[Z_{tj}^{(i)}] = 0$, $\text{E}[Z_{tj}^{(i)}*Z_{tj}^{(i)}] = p_j(1-p_j)< \infty$. Therefore by the central limit theorem for dependent variables \citep[Theorem 27.4, pg. 364]{billingsley:1995:PM} $n_i^{-1/2}\sum_{t=1}^{n_i} \mathbf{Z}^{(i)}$ is asymptotically normal with mean $\bm{0}$ and covariance matrix $\bm{\Sigma}$ with elements \begin{align*}
\sigma_{jk} &= \text{E}[ Z^{(i)}_{1,j}Z^{(i)}_{1,k}] + \sum_{t=1}^\infty \left(\text{E}[Z^{(i)}_{1,j}Z^{(i)}_{1+t,k}] + \text{E}[Z^{(i)}_{1+t,j}Z^{(i)}_{1,k}]\right)\\
&= p_j(\delta_{jk} - p_k) + 2(p_j(\delta_{jk} - p_k)) \sum_{t=1}^\infty \text{corr}(X^{(i)}_1,X^{(i)}_{1+t})\\
&= p_j(\delta_{jk} - p_k)\left(1 + 2\sum_{t=1}^\infty \text{corr}(X^{(i)}_1,X^{(i)}_{1+t})\right)\\
&= cp_j(\delta_{jk} - p_k)
\end{align*} where  $c$ is given in Equation \ref{chpt2eqc}. Thus,
\[
(x_{i1},\ldots,x_{ir}) = \dfrac{1}{\sqrt{n_i}}\sum_{t=1}^{n_i} (Z^{(i)}_{t1} / \sqrt{p_1},\ldots, Z^{(i)}_{tr} / \sqrt{p_r})
\] tends to a normal distribution with mean $\bm{0}$ and covariance $c(\mathbf{I}_{r} - \sqrt{\mathbf{p}}\sqrt{\mathbf{p}}^\intercal)$ where \[
\sqrt{\mathbf{p}} = (\sqrt{p_1}, \ldots,\sqrt{p_r})
\] Since $\mathbf{x}_i$ are independent for $i=1,\ldots,s$ we have $\mathbf{x}$ is multivariate normal with mean $\bm{0}$ and covariance matrix $c\bm{\Gamma} = c(\mathbf{I}_{rs} - \bm{\Lambda})$ where $\bm{\Lambda}$ is a block diagonal matrix where the $i$th block for $i=1,\ldots,s$ is $\sqrt{\mathbf{p}}\sqrt{\mathbf{p}}^\intercal$ 

%$\sqrt{\mathbf{p}}$ is a $rs\times 1$ vector \[
% By the argument on page 225-226 of \cite{jacobs:1978:dts2}, 
%\left(\sqrt{p_1},\ldots,\sqrt{p_1}, \ldots, \sqrt{p_r},\ldots,\sqrt{p_r} \right)^\intercal
%\]
Hence the limiting distribution of $\mathbf{y}$ is also multivariate normal with mean $\bm{0}$ and covariance matrix 
\[
(\mathbf{I}_{rs} - \mathbf{B}(\mathbf{B}^\intercal\mathbf{B})^{-1}\mathbf{B}^\intercal)c\bm{\Gamma}(\mathbf{I}_{rs} - \mathbf{B}(\mathbf{B}^\intercal\mathbf{B})^{-1}\mathbf{B}^\intercal)
\] Note that the elements of $(\mathbf{I}_r - \sqrt{\mathbf{p}}\sqrt{\mathbf{p}}^\intercal)\mathbf{B}_i$ are \[
b_{jk} = \begin{cases}
\dfrac{1}{\sqrt{p_k}}(1-\sqrt{p_j}\sqrt{p_j}) + \dfrac{\sqrt{p_j}\sqrt{p_r}}{\sqrt{p_r}} = \dfrac{1}{\sqrt{p_j}} &\text{if $j=k$ and $j< r$}\\
\dfrac{1}{\sqrt{p_k}}(-\sqrt{p_j}\sqrt{p_k}) + \dfrac{\sqrt{p_j}\sqrt{p_r}}{\sqrt{p_r}} = \sqrt{p_j} - \sqrt{p_j} = 0 &\text{if $j\neq k$ and $j < r$}\\
\dfrac{-\sqrt{p_r}\sqrt{p_k}}{\sqrt{p_k}} - \dfrac{1}{\sqrt{p_r}}(1 - \sqrt{p_r}\sqrt{p_r}) = \dfrac{-1}{\sqrt{p_r}} &\text{if $j=r$ and $k=1,\ldots,r-1$}
\end{cases}
\] Thus, $\bm{\Gamma}\mathbf{B} = \mathbf{B}$ and because $\bm{\Gamma}$ is symmetric $\mathbf{B}^\intercal\bm{\Gamma} = \mathbf{B}^\intercal$. The asymptotic covariance of $\mathbf{y}$ can then be expressed as \begin{align*}
&= (\mathbf{I}_{rs} - \mathbf{B}(\mathbf{B}^\intercal\mathbf{B})^{-1}\mathbf{B}^\intercal)c\bm{\Gamma}(\mathbf{I}_{rs} - \mathbf{B}(\mathbf{B}^\intercal\mathbf{B})^{-1}\mathbf{B}^\intercal)\\
&= c\bm{\Gamma}(\mathbf{I}_{rs} - \mathbf{B}(\mathbf{B}^\intercal\mathbf{B})^{-1}\mathbf{B}^\intercal) - c(\mathbf{I}_{rs} - \mathbf{B}(\mathbf{B}^\intercal\mathbf{B})^{-1}\mathbf{B}^\intercal)\bm{\Gamma}(\mathbf{I}_{rs} - \mathbf{B}(\mathbf{B}^\intercal\mathbf{B})^{-1}\mathbf{B}^\intercal)\\
&= c\bm{\Gamma} - c\underbrace{\bm{\Gamma}\mathbf{B}}_{\mathbf{B}}(\mathbf{B}^\intercal\mathbf{B})^{-1}\mathbf{B}^\intercal - c\mathbf{B}(\mathbf{B}^\intercal\mathbf{B})^{-1}\underbrace{\mathbf{B}^\intercal\bm{\Gamma}}_{\mathbf{B}^\intercal} + c\mathbf{B}(\mathbf{B}^\intercal\mathbf{B})^{-1}\mathbf{B}^\intercal\underbrace{\bm{\Gamma}\mathbf{B}}_{\mathbf{B}}(\mathbf{B}^\intercal\mathbf{B})^{-1}\mathbf{B}^\intercal\\
&=c\bm{\Gamma} - c\mathbf{B}(\mathbf{B}^\intercal\mathbf{B})^{-1}\mathbf{B}^\intercal - c\mathbf{B}(\mathbf{B}^\intercal\mathbf{B})^{-1}\mathbf{B}^\intercal + c\mathbf{B}(\mathbf{B}^\intercal\mathbf{B})^{-1}\mathbf{B}^\intercal\\
&= c(\bm{\Gamma} - \mathbf{B}(\mathbf{B}^\intercal\mathbf{B})^{-1}\mathbf{B}^\intercal)\\
&= c(\mathbf{I}_{rs} - \bm{\Lambda} -  \mathbf{B}(\mathbf{B}^\intercal\mathbf{B})^{-1}\mathbf{B}^\intercal)
\end{align*}
%\[
%c[\mathbf{I} - \mathbf{B}(\mathbf{B}^\intercal\mathbf{B})^{-1}\mathbf{B}^\intercal][\mathbf{I} - \sqrt{\mathbf{p}}\sqrt{\mathbf{p}}^\intercal][\mathbf{I} - \mathbf{B}(\mathbf{B}^\intercal\mathbf{B})^{-1}\mathbf{B}^\intercal] = c[\mathbf{I} - \sqrt{\mathbf{p}}\sqrt{\mathbf{p}}^\intercal - \mathbf{B}(\mathbf{B}^\intercal\mathbf{B})^{-1}\mathbf{B}^\intercal]
%\]

To obtain the asymptotic $\chi^2$ distribution of the test statistic with the appropriate degrees of freedom it is necessary to show that $\mathbf{D} = c(\bm{\Gamma} - \mathbf{B}(\mathbf{B}^\intercal\mathbf{B})^{-1}\mathbf{B}^\intercal)$ has eigenvalues 1 of multiplicity $(r-1)(s-1)$ and the rest 0. To that end, note that for an invertible matrix $\mathbf{K}$ the matrix $\mathbf{D}$ and $\mathbf{K}^{-1}\mathbf{D}\mathbf{K}$ have the same eigenvalues. Such a $\mathbf{K}$ will be constructed to obtain the eigenvalues of $\mathbf{D}$. 

The $r-1$ eigenvalues of symmetric $\mathbf{B}^\intercal\mathbf{B}$ are all positive. Denote the eigenvalues of $\mathbf{B}^\intercal\mathbf{B}$ by $\lambda_1,\ldots,\lambda_{r-1}$. By singular value decomposition (SVD) we have $\mathbf{B}^\intercal\mathbf{B} = \mathbf{C}\mathbf{M}^2\mathbf{C}^\intercal$ where $\mathbf{M}$ is a diagonal matrix with values $\sqrt{\lambda_1},\ldots,\sqrt{\lambda_{r-1}}$. Then \begin{align*}
(\mathbf{B}^\intercal\mathbf{B})^{-1} &= (\mathbf{C}\mathbf{M}^2\mathbf{C}^\intercal)^{-1} \\
&= \mathbf{C}\mathbf{M}^{-1}\mathbf{M}^{-1}\mathbf{C}^\intercal \\
\mathbf{B}(\mathbf{B}^\intercal\mathbf{B})^{-1}\mathbf{B}^\intercal &=  \mathbf{B}\mathbf{C}\mathbf{M}^{-1}\mathbf{M}^{-1}\mathbf{C}^\intercal\mathbf{B}^\intercal\\
&= \mathbf{H}\mathbf{H}^\intercal
\end{align*} where $\mathbf{H} = \mathbf{B}\mathbf{C}\mathbf{M}^{-1}$ is an $rs \times (r-1)$ matrix. Note that \[
\mathbf{H}^\intercal\mathbf{H} = \mathbf{M}^{-1}\mathbf{C}^\intercal\mathbf{B}^\intercal\mathbf{B}\mathbf{C}\mathbf{M}^{-1} = \mathbf{M}^{-1}\mathbf{M}^2\mathbf{M}^{-1} = \mathbf{I}_{r-1}
\] Thus the columns of $\mathbf{H}$ are orthonormal. Let $\mathbf{q}_i$ be a $rs\times 1$ vector where the $i$th block of $r$ variables is equal to $\sqrt{\mathbf{p}}$ i.e. \begin{align*}
\mathbf{q}_1 &= (\underbrace{\sqrt{p_1},\ldots,\sqrt{p_r}}_{r}, \underbrace{0, \ldots, 0}_{rs - r})^\intercal\\
\mathbf{q}_2 &= (\underbrace{0,\ldots,0}_{r},\underbrace{\sqrt{p_1},\ldots,\sqrt{p_r}}_{r}, \underbrace{0, \ldots, 0}_{rs - 2r})^\intercal\\
&\vdots\\
\mathbf{q}_s &= (\underbrace{0,\ldots,0}_{rs-r},\underbrace{\sqrt{p_1},\ldots,\sqrt{p_r}}_{r})^\intercal 
\end{align*} Since \[
\mathbf{B}_i^\intercal \sqrt{\mathbf{p}} = \begin{pmatrix} 
\displaystyle \sum_{j=1}^r \dfrac{\partial p_j^{(i)}}{\partial p_1} \\ \vdots \\
\displaystyle \sum_{j=1}^r \dfrac{\partial p_j^{(i)}}{\partial p_{r-1}}
\end{pmatrix} = \begin{pmatrix} 0 \\ \vdots \\ 0\end{pmatrix}
\] therefore $\mathbf{B}^\intercal\mathbf{q}_i = \bm{0}$ and $\mathbf{H}^\intercal\mathbf{q}_i = \mathbf{M}^{-1}\mathbf{C}^\intercal\mathbf{B}^\intercal\mathbf{q}_i = \bm{0}$ for $i=1,\ldots,s$. Furthermore, since $\mathbf{q}_i^\intercal\mathbf{q}_i = 1$ and $\mathbf{q}^\intercal_i\mathbf{q}_j =0$ for $i\neq j$, the $s$ vectors $\mathbf{q}_1,\ldots,\mathbf{q}_s$ can be added as columns to $\mathbf{H}$ and maintain orthonormality of $\mathbf{H}$. Let $\mathbf{H}^* = (\mathbf{H} | \mathbf{q}_1 \hdots, \mathbf{q}_s)$ be the $rs\times (s+r-1)$ matrix obtained by adding $\mathbf{q}_1,\ldots,\mathbf{q}_s$ as columns to $\mathbf{H}$. Since columns of $\mathbf{H}^*$ are orthonormal and $s+r-1 < rs$ by \cite[\S 11.9, pg. 113]{cramer:1946:mms} $rs-(s+r-1)$ columns can be added to obtain a $rs\times rs$ matrix $\mathbf{K}$ that is orthogonal. Let the last $s+r-1$ columns of $\mathbf{K}$ be equal to $\mathbf{H}^*$. 

Now, by multiplication, $\mathbf{K}^\intercal\bm{\Lambda}\mathbf{K}$ is a diagonal matrix where all values on the diagonal are 0 except for the last $s$ which are 1. Similarly, $\mathbf{K}^\intercal\mathbf{H}\mathbf{H}^\intercal\mathbf{K}$ is diagonal with diagonal values all 0 except for the $r-1$ values preceding the last $s$ values. 

Then, $\mathbf{K}^\intercal(\mathbf{I}_{rs} - \bm{\Lambda} - \mathbf{H}\mathbf{H}^\intercal)\mathbf{K}$ is a diagonal matrix which has the first $rs - s - (r-1) = (r-1)(s-1)$ values equal to 1 and the rest are 0. Therefore $c(\bm{\Gamma} - \mathbf{B}(\mathbf{B}^\intercal\mathbf{B})^{-1}\mathbf{B}^\intercal)$ has eigenvalues 1 of multiplicity $(r-1)(s-1)$ and the rest 0.

Finally, note that the test statistic of interest $X^2/c = \sum_{i=1}^s\sum_{j=1}^r y_{ij}^2/c$. Since $\mathbf{D}$ has eigenvalues 1 of multiplicity $(r-1)(s-1)$ and the rest 0, by \cite[Lemma 17.1, pg 242]{vandervaart:1998:AS} $X^2/c$ is asymptotically $\chi^2_{(r-1)(s-1)}$. 

%Now we have the test statistic $X^2/c = \sum_{i=1}^s\sum_{j=1}^r y_{ij}^2/c$, and the asymptotic $\chi^2$ distribution (with $(r-1)(s-1)$ degrees of freedom) for test statistic $X^2/c$ follows from the last part of the proof in \cite{cramer:1946:mms} on page 433. 

\pagebreak


\clearpage
\bibliographystyle{chicago}
\bibliography{ref}
\end{document}